\documentstyle[cjaa209,twoside,epsf]{article}
\input{psfig.sty}

\begin{document}

\title{GMRT Observations of Microquasar V4641 Sgr}
\author{C. H. Ishwara-Chandra and A. Pramesh Rao}
\inst{National Center for Radio Astrophysics, TIFR, P. B. No. 3,
Ganeshkhind, Pune - 7, India \\
\email{CHIC: ishwar@ncra.tifr.res.in}
}

\date{}

\abstract{We report the GMRT observations of V4641 Sgr during the May
2002 outburst at radio frequencies of 610 and 244 MHz. This is the
lowest frequency radio detection of this source. The present low
frequency radio observations clearly showed spectral evolution from the
optically thick to thin state. This behavior is broadly consistent with
the expanding bubble model. However, the flux densities observed at
lower frequencies are much higher than predicted by this model. In the
conical jet model, this discrepancy could be reconciled.

}

\section{Introduction}

The microquasar V4641 Sgr (SAX J1819.3$-$2525) was discovered by
{\tt BeppoSax} and {\tt RXTE} in 1999. Optical spectroscopy and photometry
of the source showed that V4641 Sgr is a black hole candidate in a 
binary system with a high mass companion (Orosz et al. 2001).
The object also showed 
flares in the optical (Uemura et al. 2002) and radio (Hjellming
et al. 2000). The radio flare lasted only a few days. The radio
morphology of this source suggests that it is a relativistic
jet source, similar to GRS1915+105 and GROJ 1655$-$40 (Hjellming
et al. 2000). In this paper, we present low frequency radio
observations of V4641 Sgr with GMRT at 610 and 244 MHz during
its radio outburst in May 2002.

\section{Observation and Results}

The present observations at 610 and 244 MHz were carried out with Giant
Metrewave Radio Telescope (GMRT; Swarup et al. 1991) during the radio
flare in May 2002. GMRT consists of 30 antennas, each of 45 metre
diameter spread over about 25 km south of Mumbai, India. It is the
world's largest radio telescope at metrewavelengths. Some of the useful
parameters of GMRT are given in Table 1, more details about the telescope
can be found in {\tt www.ncra.tifr.res.in}. GMRT currently has a facility
to observe simultaneously at 610 and 244 MHz, which was not operational
during these observations. However, whenever possible, the observations
at 610 and 244 MHz were taken nearly simultaneously.  The observational
details are given in Table 2. The flux density scale is set by observing
the primary calibrator 3C286 or 3C48. A phase calibrator was observed
before and after a 30 min scan on V4641 Sgr for phase calibration. The
integration time was 16 s. The data recorded from GMRT have been converted
to FITS and were analysed using Astronomical Image Processing System ({\tt
AIPS}). A few iterations of phase self calibration have been performed to
reduce the phase errors and improve the image quality. The flux density
at 610 and 244 MHz has been corrected for increased background noise in
the direction of V4641 Sgr.

\begin{center}
\begin{table}[t]
\caption{Some useful parameters of GMRT}
\hspace{0.75in}
\begin{tabular}{l l l l l l }
          &          &           &    &             &            \\
Frequency (MHz) & 151  & 235  & 325  & 610  & 1000 $-$ 1450      \\
Primary Beam ($^\circ$)& 3.8  & 2.5  & 1.8  & 0.9  & 0.56 - 0.4 \\
Resolution ($^{''}$)   &  20  & 13  &  9  &  5  & 2 - 3 \\
RMS (mJy)/hour  &   ??  & 2   & 1   & 0.5   &  0.1  \\
\end{tabular}
\end{table}
\end{center}

\begin{center}
\begin{table}
\caption{Observation log and fluxes}
\hspace{0.5in}
\begin{tabular}{l l c l l}
Date 2002 & Frequency & Duration & Flux(mJy)&   Flux(mJy)    \\
May-(UT) &  (MHz)     & (min)   &  (V4641)  &  (Bg. Source) \\
            &         &     &          &                \\
24.03       &   610   & 32  & 101.8 $\pm$5.25 &   65.2 $\pm$3.39 \\
25.02       &   610   & 25  & 102.5 $\pm$5.33 &   66.3 $\pm$3.44 \\
25.92       &   610   & 75  & 47.8 $\pm$2.97  &   71.0 $\pm$3.66 \\
26.92       &   610   & 80  & 14.4 $\pm$1.75  &   71.3 $\pm$3.68 \\
28.01       &   610   & 45  & 7.6 $\pm$ 1.38  &   67.9 $\pm$3.51 \\
29.76       &   610   & 28  & 1.4 $\pm$ 1.28  &   61.2 $\pm$3.20 \\
            &         &     &          &                \\
23.99       &   241   & 28  & 57.8 $\pm$7.70 &   102.9 $\pm$ 6.3\\
24.89       &   241   & 45  & 81.8 $\pm$5.04&   105.0 $\pm$ 8.2\\
27.85       &   241   & 30  & $< 19.9^\dagger$&   105.8 $\pm$ 10.1\\
28.90       &   241   & 30  & $< 20.3^\dagger$&   107.5 $\pm$ 13.6\\
\end{tabular}

\hspace{0.5in}
\noindent $^\dagger$Limits are 5 $\sigma$ upper limits
\end{table}
\end{center}

\section{Discussion}

This is the first positive detection of V4641 Sgr at low frequency 
of 244 MHz. Images of the field at 610 and 244 MHz are presented in 
Figure 1. At this frequency, the peak of the radio flare is 
expected at a later time as compared to higher frequencies.
The radiative lifetime of the electrons is also longer
at these frequencies, making them visible for longer duration
as compared to higher frequencies. Below we discuss the radio 
light curve and the plasma expansion model to understand 
the observed spectral change.

\begin{figure}[h]
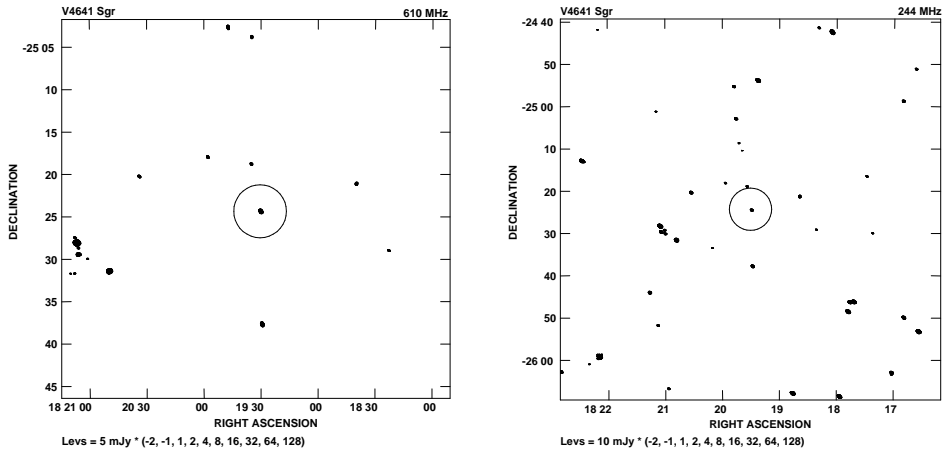

\hbox{
\hspace{0.15in}
\psfig{figure=chandra_v46_f1.ps,height=2.5in}
\hspace{0.10in}
\psfig{figure=chandra_v46_f2.ps,height=2.5in}
}
\caption{Field of V4641 Sgr at 610 (left) and 244 MHz (right).
The circle drawn at the central region is only to help to
locate the V4641 Sgr, which is at the center of this circle.}

\end{figure}

\subsection{The radio light curve}

The radio light curve of V4641 Sgr from the present observation is 
shown in Figure 2. For comparison, the flux density of a control 
source in the background is also given in the Table 2. 
At 610 MHz, the flux
density of V4641 Sgr was constant on May 24 and 25, 2002, while
the 244 MHz flux density increased from 58 to 82 mJy during the 
same period. This suggests that the source was optically thick
on May 24 and became partially optically thin on May 25. The 
flux density decayed at both frequencies subsequently. The overall
profile of the flux density decay is similar to the September 1999
radio outburst (Hjellming at al. 2000).

\subsection{Plasma expansion and the Spectral evolution}

The plot of spectral index with time clearly shows the spectral
evolution from optically thick to thin state. The spectral index $\alpha$
on May 24, 2002 is 0.60 (S$_\nu \propto \nu^\alpha$) which changed to
0.24 next day.  This can be qualitatively
understood in terms of plasma expansion in the bubble model. 
In this model, a blob of plasma is ejected from the accretion disk, 
which is initially optically thick and the low frequency radio emission 
self absorbed. As the plasma expands with time and becomes optically
thin, the flux density at lower frequencies rises and reaches its peak
emission. Application of the synchrotron bubble with the parameters
used by Hjellming et al. (2000) predicts the 610 MHz emission
to peak $\sim$ 37 hours after the ejection. From the ASM lightcurve
(Figure 3),
it appears that the X-ray flare occurred on May 21, 2002. If this is 
taken as the time of ejection, the 610 MHz emission should have peaked between 
May 22 and 23, 2002. This is confirmed by the peak observed by 
Molonglo Synthesis Telescope (MOST) at 843 MHz near this time (Champbell-Wilson,private communication). However, the naive application of the 
synchrotron bubble model predicts the flux density at lower
frequencies to be much less than that at higher frequencies,
which is inconsistent with the observations. This inconsistency
may be reconciled in the conical jet model (Hjellming and Johnston, 1988).

\begin{figure}
\hbox{
\hspace{0.20in}
\psfig{figure=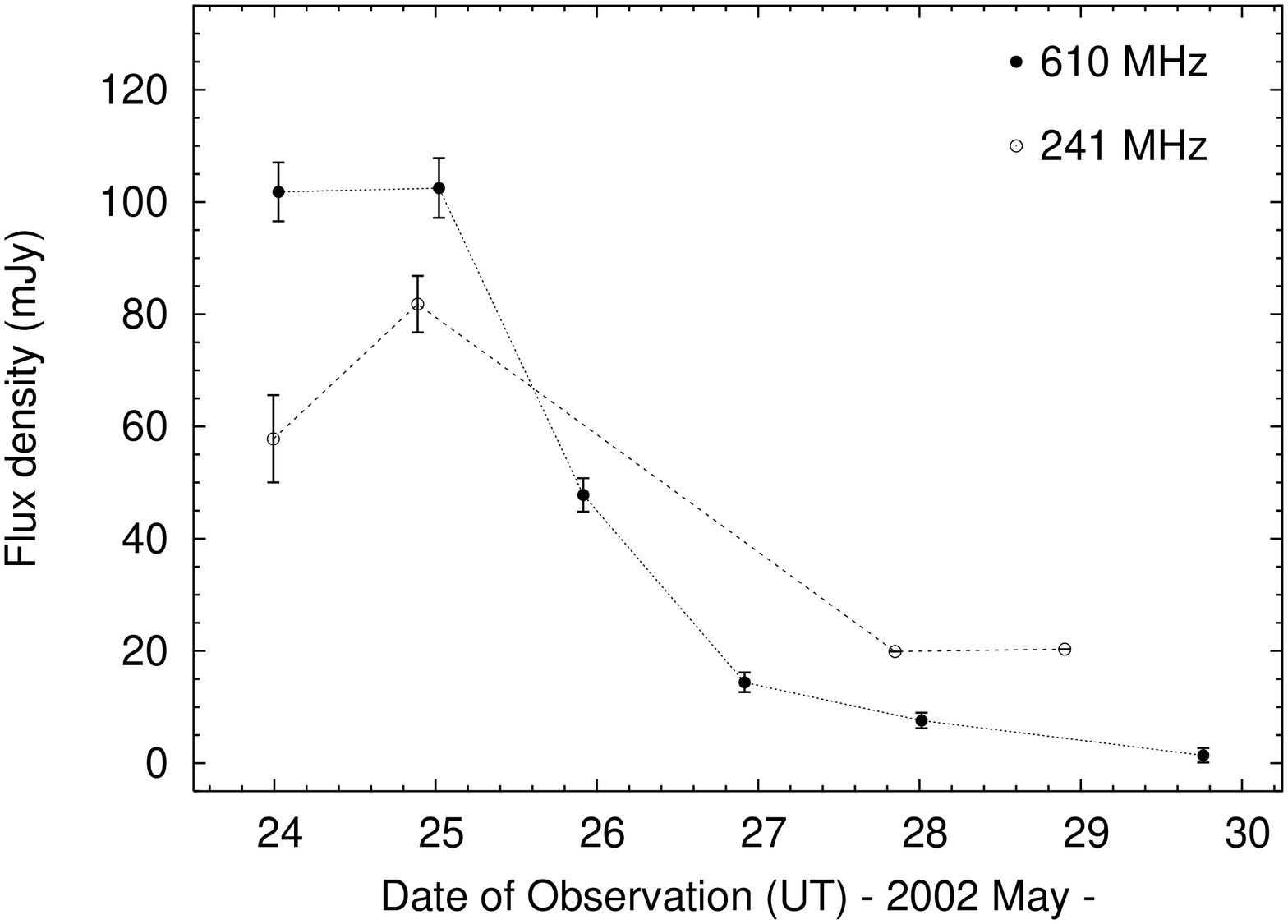,height=1.8in,angle=270}
\hspace{0.10in}
\psfig{figure=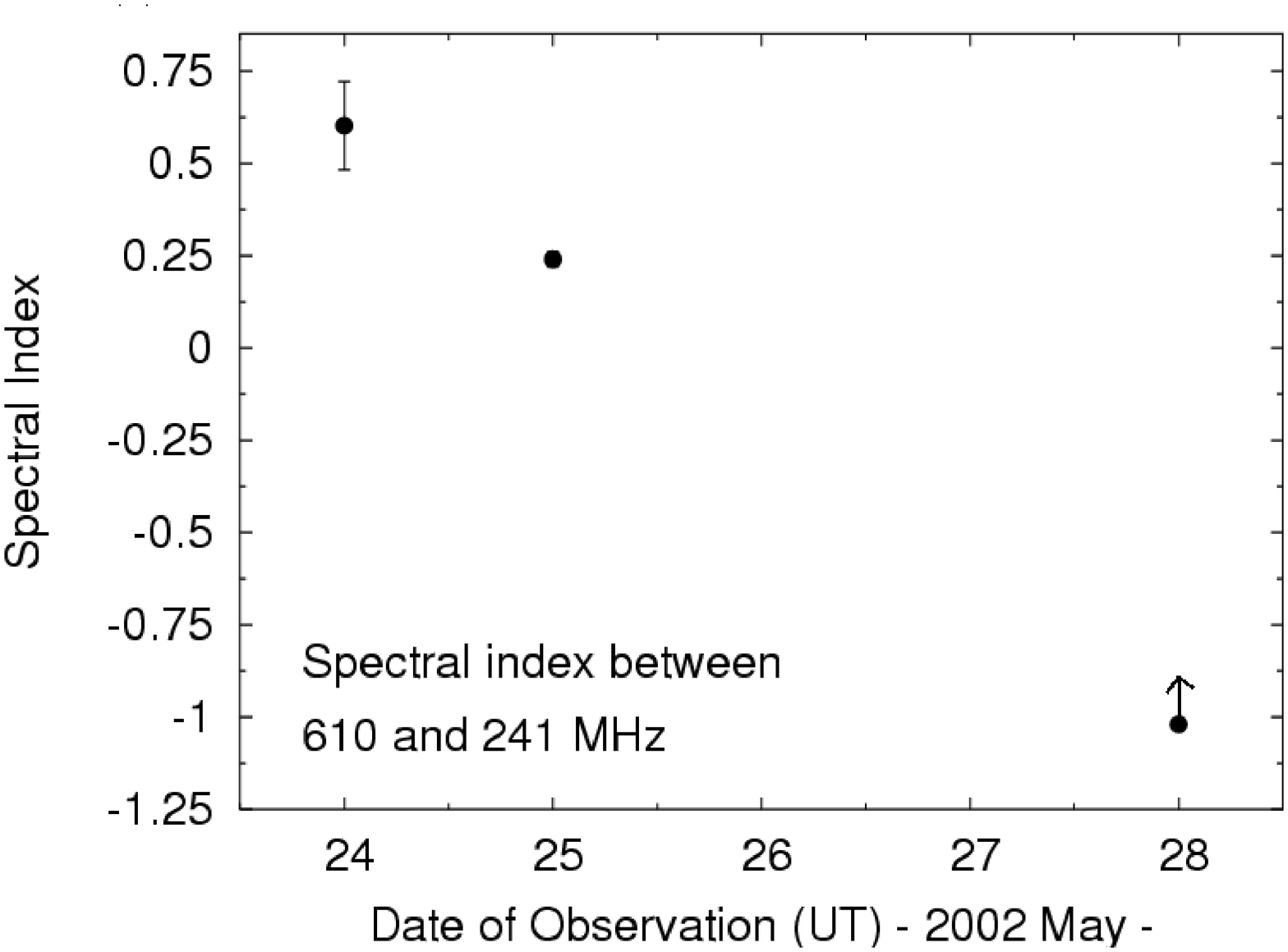,height=1.8in,angle=270}
}
\caption{Light curve of V4641 Sgr at 610 and 244 MHz (left)
and the plot of spectral index for each day with nearly 
simultaneous observation at these frequencies (right).}

\end{figure}

\begin{figure}
\hspace{0.25in}
\psfig{figure=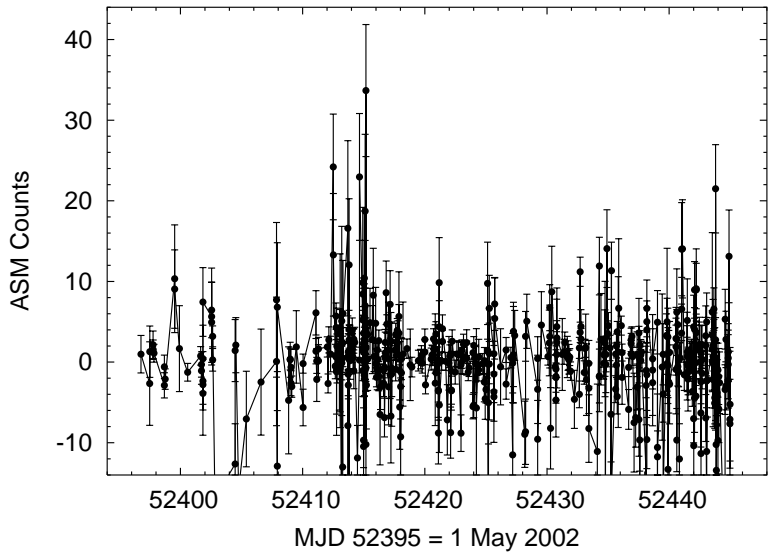,height=3.0in,angle=270}
\caption{RXTE/ASM Lightcurve for V4641 Sgr during May - June 2002.}
\end{figure}

\section{Concluding Remarks}

We have shown for the first time that the microquasar V4641 Sgr exhibits
strong radio emission at metrewavelengths. Although the time-delays
estimated by assuming the expanding bubble model agrees reasonably
with the observations,the flux density measurements at low frequencies
are significantly higher than predicted by this model. This may be in
better agreement with the conical jet model. This observational result
demonstrates the importance of the low frequency radio observations for
constraining the models of radio emission from microquasars.

\section*{Acknowledgments}

GMRT is run by the National Centre for Radio Astrophysics of the Tata
Institute of Fundamental Research. This research has made use of NASA's
Astrophysics Data System and of the SIMBAD database, operated at CDS,
Strasbourg, France.


\begin{thebibliography}{}

\bibitem[]{} Hjellming, R. M., Johnston, K. J., 1988, ApJ, 328, 600
\bibitem[]{} Hjellming, R. M., et al., 2000, ApJ, 544, 977
\bibitem[]{} Orosz, J., Kuulkers, E., van der Klis, M., et al. 2001, 
             ApJ, 555, 489
\bibitem[]{} Uemura M., Kato, T., Watanabe, T., Stubbings, R., Monard, B., 
             Kawai, N., 2002, PASJ, 54, 95
\end{thebibliography}
\end{document}